\documentclass[runningheads]{llncs}
\usepackage{graphicx}
\usepackage{algorithmic}
\usepackage{amsmath}
\usepackage[ruled,vlined]{algorithm2e}
\usepackage{multirow}
\usepackage{multicol}
\usepackage[T1]{fontenc}
\usepackage[utf8]{inputenc}
\usepackage{babel}
\usepackage[font=small,labelfont=bf]{caption}

\usepackage[printwatermark]{xwatermark}
\usepackage{xcolor}
\usepackage{graphicx}
\usepackage{lipsum}

\newwatermark[allpages,color=gray!50,angle=45,scale=1,xpos=0,ypos=0]{ACCEPTED AND PRESENTED\\at SBP-BRiMS 2021}

\begin{document}
\title{Predicting Users' Value Changes by the Friends' Influence from Social Media Usage}
%
%
\author{Md. Saddam Hossain Mukta\inst{1} \and
Ahmed Shahriar Sakib\inst{2} \and
Md. Adnanul Islam\inst{3} \and
Mohiuddin Ahmed\inst{4} \and
Mumshad Ahamed Rifat\inst{1}
}
\authorrunning{Mukta et al.}
\titlerunning{Value Changes by Friends' Influence}
%
\institute{United International University, Dhaka, Bangladesh \and
American International University Bangladesh \and
Military Institute of Science and Technology \and
Edith Cowan University, Australia 
}


%
\maketitle              
\begin{abstract}
Basic human values represent a set of values such as security, independence, success, kindness, and pleasure, which we deem important to our lives. Each of us holds different values with different degrees of significance. Existing studies show that values of a person can be identified from their social network usage. However, the value priority of a person may change over time due to different factors such as life experiences, influence, social structure and technology. Existing studies do not conduct any analysis regarding the change of users' value from the social influence, i.e., group persuasion, form the social media usage. In our research, first, we predict users' value score by the influence of friends from their social media usage. We propose a Bounded Confidence Model (BCM) based value dynamics model from 275 different ego networks in Facebook that predicts how social influence may persuade a person to change their value over time. Then, to predict better, we use particle swarm optimization based hyperparameter tuning technique. We observe that these optimized hyperparameters produce accurate future value score. We also run our approach with different machine learning based methods and find support vector regression (SVR) outperforms other regressor models. By using SVR with the best hyperparameters of BCM model, we find the lowest Mean Squared Error (MSE) score $0.00347$.

\end{abstract}
\section{Introduction}
In recent times, Social Networking Sites (SNS) have become a major platform of communications among users on the web. 
These SNS data provide a wide range of opportunities to identify cognitive and psychological attributes such as basic human values (aka \textit{values})~\cite{chen2014understanding}, personality~\cite{golbeck2011predicting}, and behavior~\cite{nawshin2020modeling}. Values represent one’s attitudes, opinions, thoughts, and goals in life. Values of an individual might amend over time due to the influence of her group of friends~\cite{danis2011values,friedkin1990social}. In this paper, we predict users' future value scores, which might be changed by the influence of friends in an egocentric network such as Facebook
\footnote{Facebook is an example of an egocentric network~\cite{arnaboldi2013egocentric} because the network provides interaction capability only among the friends while prevent any interactions from the external users to this network.}.

Values represent a set of criteria such as security, self-enhancement, etc., that are used by individuals to take different actions. Chen et al.~\cite{chen2014understanding} predict five higher-level values from user word usages in Reddit. In another study~\cite{mukta2016user}, authors predict users' values from generated (i.e., statuses) and supported (i.e., likes and shares) contents in Facebook. 
However, these approaches largely fail to capture the change of users' value priorities from the social network usage. Several socio-psychological studies~\cite{crandall2008feedback,friedkin1990social} show that values of a user might be reshaped by the influence of other members in the same interest group. To the best of our knowledge, no study has been conducted to identify users' change of value scores by the influence of friends from the social network usage. However, identifying the change of values of an individual from friends' influence has a number of applications such as identifying university course major or career path shifting trends, prediction of customers' purchasing behavior, transforming customers' product selection preference or marketing policies, and understanding transition of economics and business.  \\

To this context, we propose a technique to identify the changes of users' value from their social network interactions by using Bounded Confidence Model (BCM)~\cite{quattrociocchi2014opinion,deffuant2004modelling,gomez2012bounded}. 
Motivated by the work~\cite{chen2014understanding}, first we collect data of 275 user networks from Facebook by using a google survey form. Since users' change of values are observed in terms of time intervals, we separate users' Facebook statuses, comments, and shares according to a time span of six months~\cite{mukta2019temporal}. Then, we generate value scores from users' each six months interval Facebook statuses, comments, and shares by using IBM Watson Personality insight API~\footnote{https://www.ibm.com/watson/services/personality-insights/}. 
From the computed value scores of peer friends, we compute the hyperparameters (convergent factor, $\mu$ and threshold, $\sigma$) for our BCM model (see Equation~\ref{eq:1}). Then, we use particle swarm optimization (PSO)~\cite{bansal2019particle} method for finding the best hyperparameter configuration. Finally, we use these optimized hyperparameters of the BCM model to predict next value score by using support vector regressor (SVR) model~\cite{awad2015support}. 
In summary, our contributions in this paper can be highlighted as follows:
\begin{itemize}
	\item We demonstrate the change of values by the influence of group of friends using BCM model.
	\item We develop a PSO based best hyperparameter selection method that predicts user's future value score with less MSE score.
\end{itemize}

\section{Preliminaries and Related Work}
In this section, we describe the preliminaries about values, BCM model, and how influence may reshape users' behavior from social media usage.

\subsection{Values}
Basic Human Values define the goal, belief and behavior of an individual. Schwartz et al.~\cite{schwartz2003proposal} categorize the value dimension into five higher-level values.  \textbf{Openness-to-change} mainly refers to the ability to ``think outside of the box” which consists of two broad personal values: \textit{self-direction} and \textit{stimulation}~\cite{schwartz1992universals}. 
\textbf{Self-transcendence} satisfies the motivational goal for preservation and enhancement of the welfare of people with whom one is in frequent social contact~\cite{schwartz1992universals}.
\textbf{Self-enhancement} refers economic well-being and notoriety, control or strength over individuals and resources.
\textbf{Conservation} emphasizes order, self-restriction, preservation of the past, and resistance to change. 
\textbf{Hedonism} basically means pleasure or sensuous gratification for oneself~\cite{schwartz1992universals}.

\subsection{Bounded Confidence Model (BCM)}
BCM is a popular opinion dynamics model to determine the influence of a network of people over an individual. We use the BCM devised by Deffuant et al.~\cite{deffuant2000mixing} to understand the nature of such changes which can be different from one to another. Considering the distance of the corresponding values between two users is less than a given threshold $\sigma$, the updated value of each of the users can be calculated using the following equation:

{\small
	\begin{equation}\label{eq:1}
		BHV\textsuperscript{t+1}\textsubscript{i}=BHV\textsuperscript{t}\textsubscript{i}+ \mu\textsubscript{ego}(BHV\textsuperscript{t}\textsubscript{j}-BHV\textsuperscript{t}\textsubscript{i})\Theta(\sigma\textsubscript{ego}-|BHV\textsuperscript{t}\textsubscript{j}-BHV\textsuperscript{t}\textsubscript{i}|)
	\end{equation}
}
where BHV\textsubscript{i} is the value score of the user \textit{i}, $\mu$\textsubscript{ego} is a convergence factor, $\sigma$\textsubscript{ego} is the threshold within which the users interact or adapt with each other, and $\Theta$() is a Heaviside's theta function~\footnote{\url{https://mathworld.wolfram.com/HeavisideStepFunction.html}}. 


\subsection{Values in Social Media}
Chen et al.~\cite{chen2014understanding} identify values from Reddit, an online news sharing community. The authors identify five higher-level values from user's pattern of word using in social media. They predict the value scores by using linear regression. 
Boyd et al. identify values~\cite{boyd2015values} from statuses of 767 Facebook users. They identify values with a data driven approach. Mukta et al.~\cite{mukta2016user} identify values from both user generated and supported contents in Facebook. Mukta et al.~\cite{mukta2019temporal} predict the temporal change of values of Facebook users by using different ML techniques. In this paper, we devise a novel technique to identify the value change of an individual by the social influence of the \textit{friends} in Facebook.

\subsection{Value Changing Influence Models}
Value of a user might be changed her values for different offline behaviors such as life experiences, life events, technological change, social structure, and life style of others~\cite{danis2011values}~\cite{friedkin1999social}~\cite{danis2011values}. In a previous study~\cite{mukta2019temporal}, we find that value change of an individual reflects in their social media usage behavior. Authors capture the opinion change over time from social media. Quattrociocchi et al.~\cite{quattrociocchi2014opinion} show that inner dynamics of information systems, i.e., TV, newspaper, social network platforms, - play a vital role on the evolution of the public network.  

In this paper, we compute value scores of different time intervals and analyze the interactions among users. Then we measure how the value of one person may influence the change of value on others through social media interactions.

\section{Methodology}
\begin{figure}[hbt!]
    \centering
    \includegraphics[scale=.45]{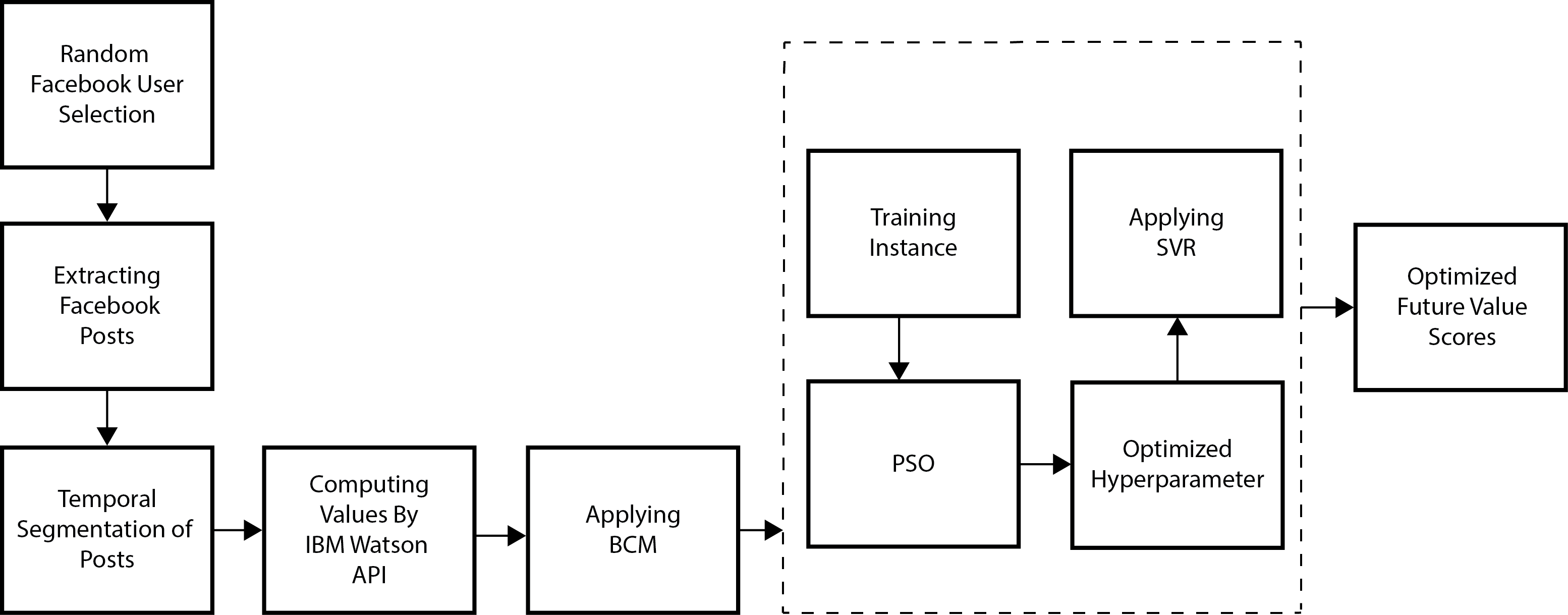}
    \caption{A complete pipeline of value change modeling and BCM hyperparameter optimization.}
    \label{fig:meth}
\end{figure}

In this section, we first describe the process of value change modeling by the influence of close friends using BCM model. Figure~\ref{fig:meth} shows the complete pipeline of our value change modeling and its hyperparameter optimization process. We discuss our methodology in the subsequent sections.

\subsection{Data Collection}
We randomly select a total of 275 (motivated by the study of Golbeck~\cite{golbeck2011predicting}) different Facebook ego networks where each user holds an ego network and his Facebook friends are alters. Initially we invite a total of 320 Facebook friends to share their ego networks, but only a total of 275 users show their interest to share their data. After selection we extract the list of users; then collect every comment of each user. Users' collect their data from their own profile~\footnote{https://www.facebook.com/settings?tab=your\_facebook\_information} and download her data~\footnote{https://www.facebook.com/dyi/?referrer=yfi\_settings} in different time intervals with different formats: HTML and Json. Later, they share their data with us through the google form.

We extract public profile data like statuses, posts and shares that support comments and likes. Then, we divide the data into temporal segments. Each year is divided by 2 segments, each segment contains data of 6 months. In our experiment, we have Facebook posts of maximum 10 years for a single user which we can divide into 20 segments. 
We select users based on their number of their daily public interactions. 
Table~\ref{stat} shows the statistics of our dataset.

\begin{table}[]
\caption{Statistics of our dataset}
\centering
\begin{tabular}{|l|r|lll}
\cline{1-2}
\textbf{Attributes}                  & \textbf{Values} &  &  &  \\ \cline{1-2}
Number of ego Networks                      & 275             &  &  &  \\ \cline{1-2}
Number of total Comments             & 75,625         &  &  &  \\ \cline{1-2}
Number of Maximum Comments of a User & 237             &  &  &  \\ \cline{1-2}
Number of Minimum Comments of a User & 25             &  &  &  \\ \cline{1-2}
Total Time Duration (years) & 10             &  &  &  \\ \cline{1-2}
Maximum Time Span for a user (years) &    10          &  &  &  \\ \cline{1-2}
Minimum Time Span for a user (years) & 7             &  &  &  \\ \cline{1-2}
\end{tabular}
\label{stat}
\end{table}


\subsection{Influence Modeling for Value Change}
Next, we first propose a new model to investigate the change of values over time. During applying the BCM model~\cite{deffuant2000mixing}, we use two hyperparameters: \textit{convergence factor ($\mu$)} and \textit{threshold ($\sigma$)} of value difference. To optimize these hyperparameters, we use a machine learning based approach where we find optimum values (i.e., solution) by using PSO algorithm (Figure~\ref{fig:meth}). From these optimized hyperparameters, we find the \textit{threshold ($\sigma$)} value for predicting the next value score through using SVR.

\section{Experimental Evaluation}
\subsection{Ego Network Configuration}
We consider the adjustment of the value score of an ego (\textit{u\textsubscript{i}}) by interacting all users \textit{u\textsubscript{j}}. We collect interaction data such as likes, comments, and sharing of object between an ego u, and their friends, \textit{u\textsubscript{j}} where \textit{j$\leq$5} according to \textit{Dunbar number}~\cite{arnaboldi2013egocentric}. We are interested in showing that when a number of users \textit{u\textsubscript{j}} influence to value score of a single user (e.g. an ego), his/her value score converges to a single unified score. This idea is borrowed from \textit{mean field theory} used in the research paper by Boudec et al.~\cite{DBLP:conf/qest/BoudecMM07}.

\begin{figure*}[t!]
    \centering
        \centering
        \includegraphics[scale=0.47]{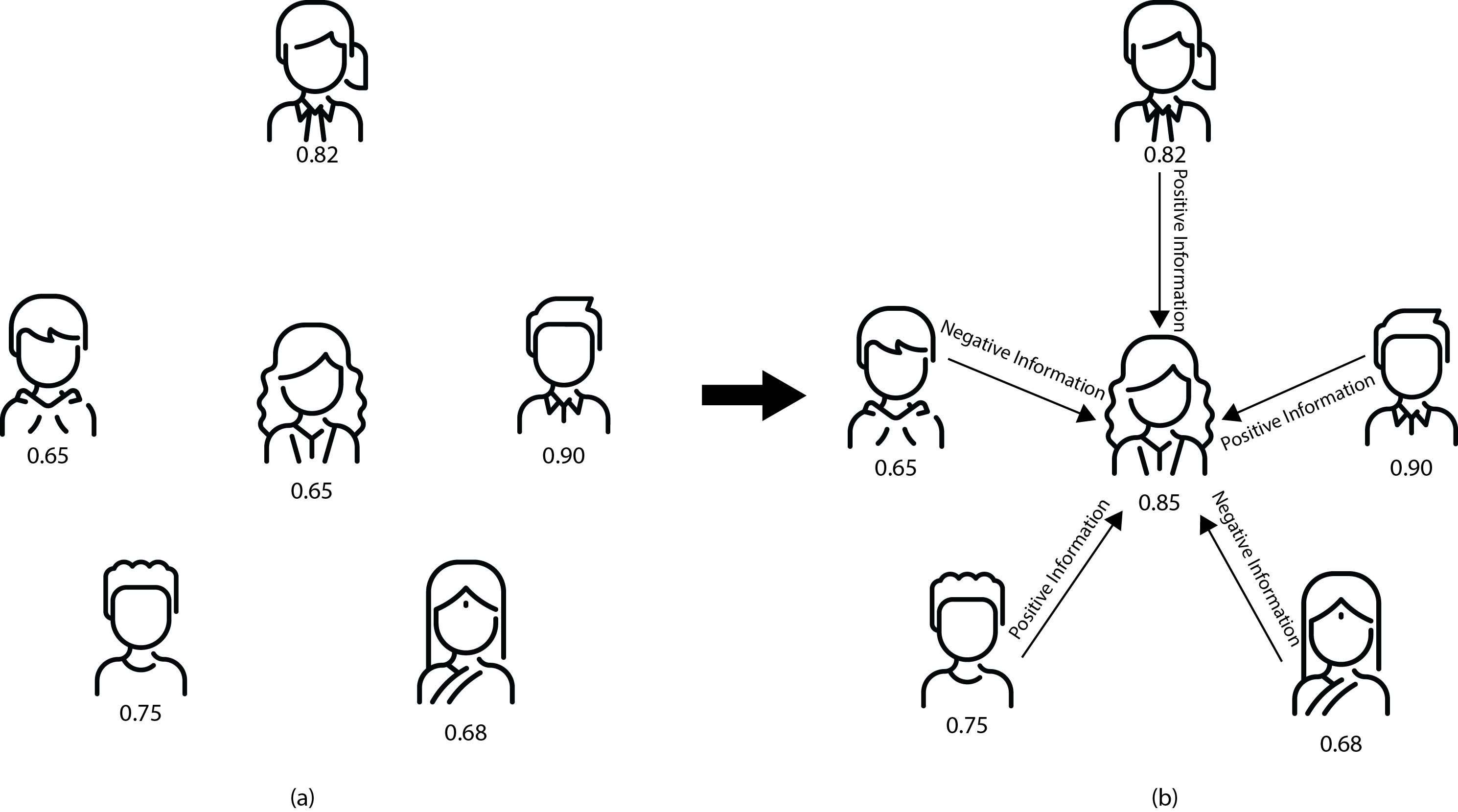}
        \caption{Group wise influence based value change from time \textit{t\textsubscript{1}} (Fig a) to \textit{t\textsubscript{2}} (Fig b)}
        \label{beforechange}
    %
    ~ 
\end{figure*}

Ego interacts with a total of five close alters (according to Dunbar number of close friends)~\cite{arnaboldi2013egocentric}. In this study, we assume that users may shift their value scores by the influence of close friends who fall in the sphere where N=5 ~\cite{arnaboldi2013egocentric}. 

Like a group discussion, we select a primary user to detect his/her value change from the influence of other people from the group. Figure~\ref{beforechange}(a) shows that independent value scores for a total of 5 Facebook users. When they interact each other in a group, value score for a user might be changed by the influence of other alters, i.e., Facebook friends. Figure~\ref{beforechange}(b)  shows the final score for a value dimension after being influenced by the members of the group. We predict the change of users value by using BCM model. The model has different hyperparameters  that we optimize by using Particle Swarm Optimization (PSO) method in subsection~\ref{he}.

\subsection{Hyperparameter Estimation}
\label{he}
The convergence factor, $\mu$ indicates the momentum term of the influence dynamics. In our study, we consider $\mu$= 0.4 following the study of~\cite{quattrociocchi2014opinion}. However, to obtain the appropriate value of the threshold ($\sigma$) for the BCM model, we build a regression model. The regression model actually predicts the $\sigma$ which minimize the   To build the regression model, we consider four features for the training instances. The features are: \romannumeral 1) value of user, \textit{u\textsubscript{i}}, at t time, \romannumeral 2) value of user's friend, \textit{u\textsubscript{j}}, at t time, \romannumeral 3) value of user, \textit{u\textsubscript{i}}, at \textit{t+1} time, and \romannumeral 4) Convergence Factor, $\mu$. 
We perform several regression models with these features by a 10-fold cross-validation with 10 iterations. We use the following regressors: SVR~\footnote{https://bit.ly/2OOBDZa}, Gaussian Process Regressor~\footnote{https://bit.ly/2OWxAKy}, ElasticNet~\footnote{https://bit.ly/3pDuh7M}, BayesianRidge~\footnote{https://bit.ly/3qJoyys} and MLPRegressor~\footnote{https://bit.ly/3qEivez}. 

For hyperparameter tuning, we apply PSO~\footnote{https://bit.ly/3ucqcLf} by using Optunity Library~\footnote{https://homes.esat.kuleuven.be/~claesenm/optunity/}. Particle swarm optimisation (PSO)~\cite{bansal2019particle} is a typical algorithm of the swarm intelligence family. The algorithm 
is a population-based meta-heuristic optimization technique which initializes a number of individual search ‘particles’, each representing a possible solution. 
This population of particles change their positions by an evolutionary process. Each of these particles is in movement with a velocity allowing them to update their position over the iterations to find the global minimum. 

\begin{table}[htb!]
\caption{Hyperparameters Configuration for different Regressors }
\label{hyperpameterConfiguration}
\begin{tabular}{|c|l|}
\hline
Regression Model               & \multicolumn{1}{c|}{Hyperparameter Configuration}                                    \\ \hline
\multirow{3}{*}{SVR}           & kernel -RBF, gamma- {[}0, 50{]} , C-{[}1, 100{]} ;                                   \\ \cline{2-2} 
                               & kernel -linear,C-{[}1, 100{]} ;                                                      \\ \cline{2-2} 
                               & kernel -poly, degree- {[}2, 5{]} , C-{[}1000, 20000{]}, coef0-{[}0,1{]}              \\ \hline
Gaussian Processes             & normalize\_y= {[}True,False{]}, alpha = {[}1$e$-10 - 1$e$-2{]}                         \\ \hline
ElasticNet                     & alpha - {[}0, 1.0{]} , l1\_ratio - {[}0, 1.0{]} , tol - {[}1$e$-4, 0.01{]}            \\ \hline
\multirow{2}{*}{BayesianRidge} & alpha\_1 - {[}1$e$-6, 0.01{]}, alpha\_2 - {[}1$e$-6, 0.01{]},                          \\ \cline{2-2} 
                               & lambda\_1 - {[}1$e$-6, 0.01{]}, lambda\_2 = {[}1$e$-6, 0.01{]}, tol -{[}1$e$-4, 0.01{]} \\ \hline
\multirow{2}{*}{MLPRegressor}  & hidden\_layer\_sizes - {[}( 50, 50, 50), ( 50, 100, 50 ), ( 100, ){]} ,              \\ \cline{2-2} 
                               & activation - {[}’tanh’, ’relu’{]}, alpha - {[}1$e$-4, 0.01{]}                         \\ \hline
\end{tabular}
\end{table}

\begin{table}[!htbp]
\caption{Strength (Low RMSE) of the regression model }
\centering
\begin{tabular}{|l|r|lll}
\cline{1-2}
\textbf{Regression Algorithm}                  & \textbf{MSE} &  &  &  \\ \cline{1-2}
SVR                     & 0.00334             &  &  &  \\ \cline{1-2}
Gaussian Process Regressor           &  0.07138           &  &  &  \\ \cline{1-2}
MLPRegressor & 0.08912             &  &  &  \\ \cline{1-2}
ElasticNet & 0.05312             &  &  &  \\ \cline{1-2}
BayesianRidge & 0.01183         &  &  &  \\ \cline{1-2}
\end{tabular}
\label{strength}
\end{table}

ParticleSwarm~\footnote{https://bit.ly/37zcGYb} has 5 parameters that can be configured: \textit{num\_particles, num\_generations, $\phi$\textsubscript{1}, $\phi$\textsubscript{2},} and \textit{max\_speed}. In our experiment, we use \textit{num\_particles} = 10 , \textit{num\_generations} = 15 ,\textit{max\_speed } = None, $\phi$\textsubscript{1} = 1.5 , $\phi$\textsubscript{2} = 2.0 to initialize ParticleSwarm~\footnote{https://bit.ly/3aBdSw3} solver in \textit{Optunity}. In our case, we consider one set of hyperparameter configuration as a single particle. 

We split the dataset into training, test, and validation data by 70\%, 20\% and 10\%, respectively. Then, we define our objective function to minimize the cost of the model which in this case, Mean Squared Error (MSE). We initialize different box constrained configuration sets of hyperparameters for different regressors. Each particle represents a configuration for hyperparameters of the Machine Learning Model. All of the particles have MSE, i.e. fitness value which are evaluated by the cost function to be minimized. The particles move through the problem space by following the current optimum particles. 
Since \textit{Optunity} can optimize conditional search spaces , we set different hyparameters based on the kernel. We use \textit{Radial Basis Function (RBF)}, \textit{linear} and \textit{polynomial } kernel. Table~\ref{hyperpameterConfiguration} shows the search space for \textit{SVR} hyperparameters. Table~\ref{hyperpameterConfiguration} shows the optimized hyperparameter configuration values for the regressors. Table \ref{strength} shows the performance of the regressors. We find SVR regressor shows the best average performance (MSE-0.00337) to predict users’ threshold value on the BCM model. Algorithm~\ref{hypoptimize} presents the process of our PSO based hyperparameter optimization method.


\begin{algorithm}[tbh]
\begin{multicols}{2}
\SetKwInOut{Input} {initialize}  
\Input{\\
 $u\textsubscript{i}$ as value of a user\\
 $u\textsubscript{j\dots n}$ as values of group users \\
 $u\textsubscript{i}+t$ as value of a user in time \textit{t} \\
 $\mu$ as convergence factor ( 0.4 ) \\
 $\sigma$ as Threshold Value 

}
\textbf{Proc.: \textit{Dataset\_Prep\_BCM( )}}\\
1: Calculate the threshold value via BCM Model \\
2: Set predictor variables ($X$) :  $u\textsubscript{i}$,  $u\textsubscript{j\dots n}$ , $u\textsubscript{i}+t$ , $\mu$\\
3: Set Dependent variable :  $\sigma$\\
4: Divide the Data set by 70\% and 30\% as D1, and
D2, Respectively \\

\textbf{Proc.: \textit{Model\_Training( )}}\\
5: initialize D1 as train dataset\\
6: initialize D2 as test dataset\\
7: Run ML models to predict $\sigma$\\

\textbf{Proc.: \textit{Hyperparam\_Optim( )}}\\
\textbf{PSO: Initialize}\\
    num\_particles=10 \\
    num\_generations=15 \\
    max\_speed=None \\
    $\phi$\textsubscript{1}=1.5 \\
    $\phi$\textsubscript{2}=2.0\\

8: Find the optimum $\sigma$ for BCM model using PSO \\
9: Save ML model as M1\\
10: Calculate the $u\textsubscript{i}+t$ using M1\\
11: Calculate MSE on D2 dataset with respect to M1
\caption{PSO\_based\_BCM\_Hyperpar\_Optimization}
\label{hypoptimize}
\end{multicols}
\end{algorithm}

\section{Results and Discussion}




\begin{figure}[htbp]
    \centering
    \includegraphics[scale=0.3]{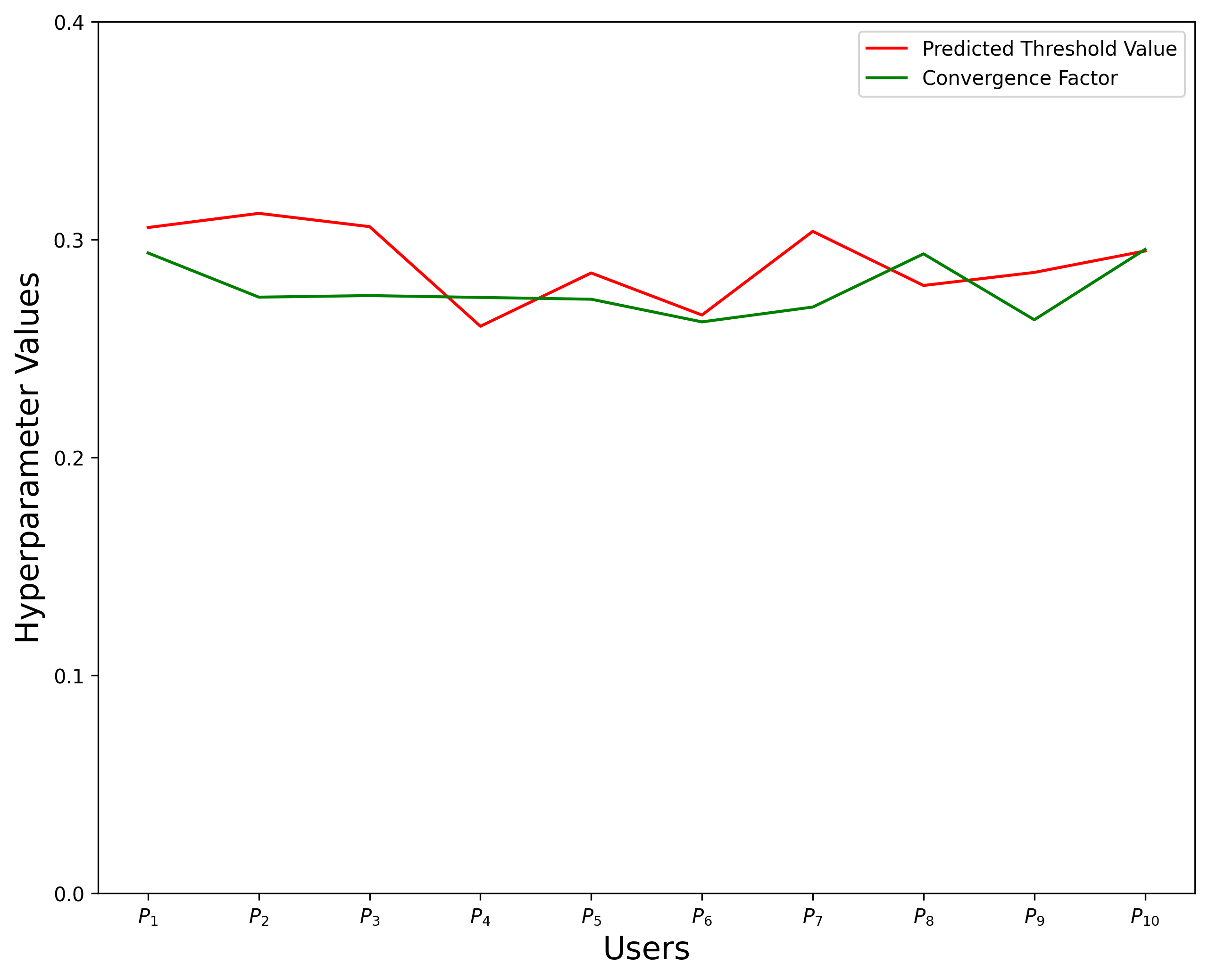}
    \caption{Users' value Scores computation and its hyperparameters.}
    \label{fig3}
\end{figure}

\begin{figure}[ht]
    \begin{minipage}[b]{0.45\linewidth}
    \centering
    \includegraphics[width=\textwidth]{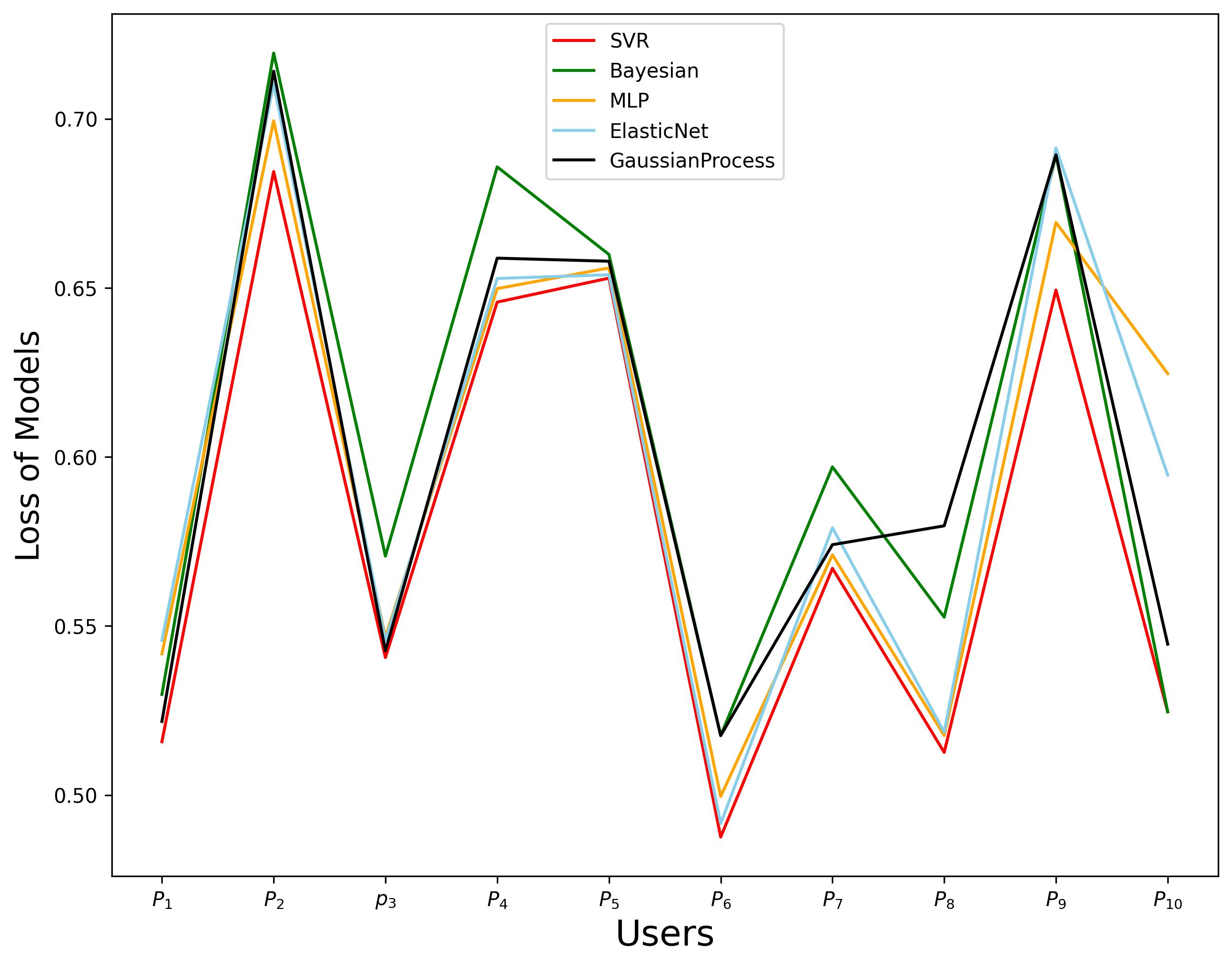}
    \caption{Loss of users' predicted value scores by SVR.}
    \label{fig4}
    \end{minipage}
    \hspace{0.5cm}
    \begin{minipage}[b]{0.45\linewidth}
    \centering
    \includegraphics[width=\textwidth]{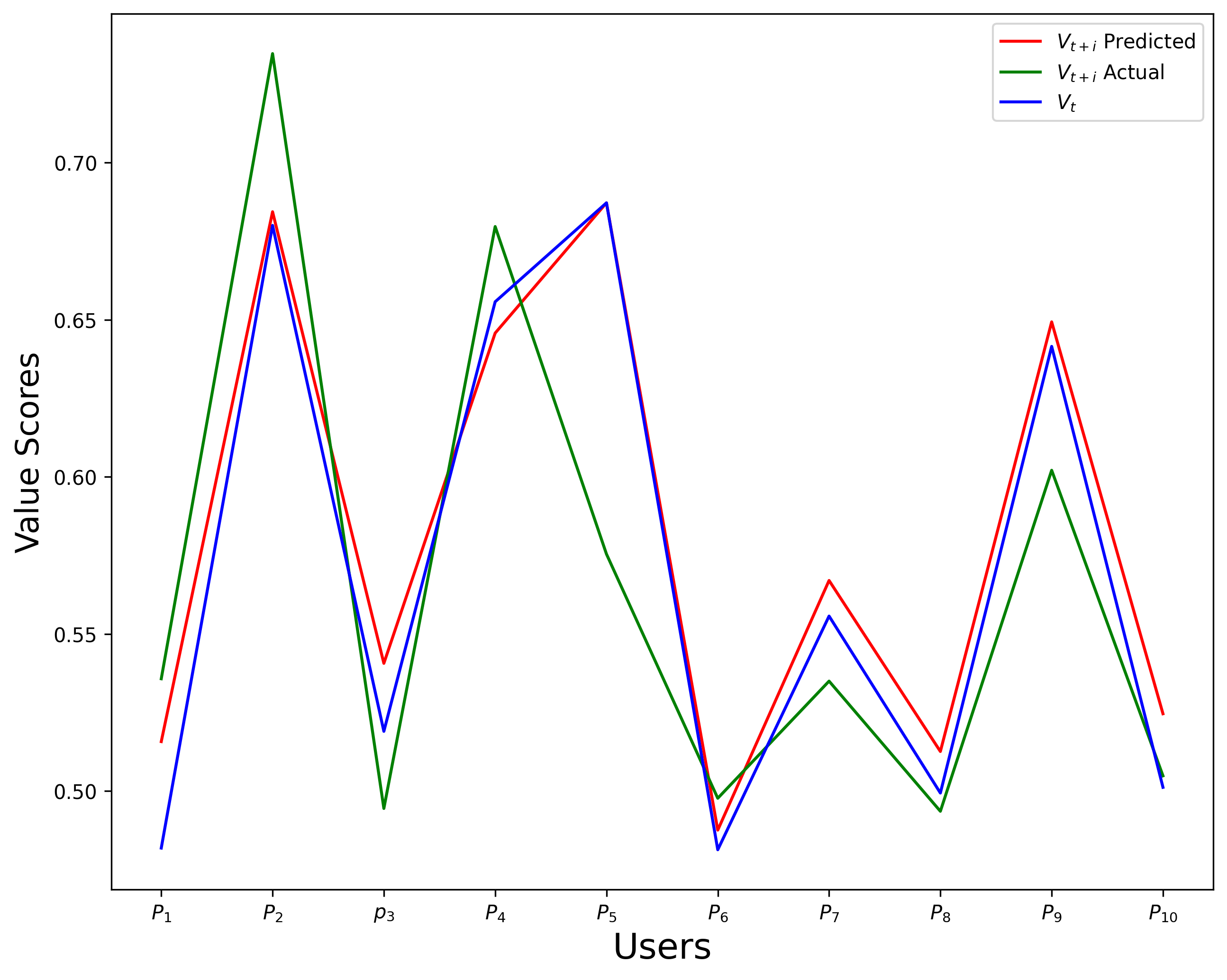}
    \caption{Users' Value scores by using the best hyperparameters.}
    \label{fig5}
    \end{minipage}
\end{figure}

In this experiment, we take Facebook interactions (i.e., status, comments, shares, and likes) of 275 users. Figure~\ref{fig3} presents the variation of convergent factor and threshold for computing values by using BCM models. By using PSO, we find the best set of parameters for computing SVR. With these parameters, we predict the best $\sigma$ by using SVR. Figure~\ref{fig4} presents the loss to compute the $\sigma$ by using different models. Among these models, SVR shows the lowest loss to predict the hyperparameters. When we use the $\sigma$ value, we predict the best final values scores of a user by the influence her close friends. Figure~\ref{fig5} shows the actual future value scores and predicted value scores.      


Bardi et al.~\cite{bardi2009structure} show that wide range of influence can change one's values across contexts and time. Rokeach et al.~\cite{rokeach1973nature} observe that in addition to cultural and societal changes, personal values  might be changed. The author also observe that values likely to change during adulthood. Roberts et al.~\cite{roberts2003cumulative} personality and values are subject to change and adaption across different life stages.

Berndt~\cite{berndt1992friendship} describes that friendships have influence on user's attitude and behavior. For example, adolescents whose friends drink beer at parties likely to start drinking. In contrast, user's value may influence negatively. For example, adolescents often have conflicts often have conflicts with others which might propagate among others. Epstein~\cite{epstein1983influence} and Hartup~\cite{hartup1993adolescents} describe that friends influence each other in different behaviors, including aspirations, achievements, values and attitudes, social skills, and appropriate sex roles. 

\section{Conclusion}
In this paper, we have extracted 275 different ego networks from a Facebook. Then, we have identified intimate friends for each of the ego networks by using Dunbar number. Then, we have segmented users' interaction in a time frame of 6 months. Then, we have computed users' value scores from their Facebook interactions by using IBM personality insight API. Based on the users' value scores, we have proposed a value dynamic technique based on BCM influence model. During modeling, we have also proposed a PSO based hyperparameter estimation technique. Our model have showed an outstanding performance (i.e., lower MSE) in predicting change of users' value from their social media interactions.


\bibliographystyle{splncs04}
\bibliography{ref_bibliography}
%




\end{document}